\documentclass[pra,twocolumn,showpacs,amsmath,amssymb]{revtex4}
\usepackage{graphicx}
\usepackage{dcolumn}
\usepackage{bm}

\begin{document}

\title{Quantum Computation with Diatomic Bits in Optical Lattices}

\author{Chaohong  Lee}
 \altaffiliation{E-mail: chl124@rsphysse.anu.edu.au; chleecn@gmail.com}

\author{Elena A. Ostrovskaya}

\address{Nonlinear Physics Centre and ARC Centre of Excellence for
Quantum-Atom Optics, Research School of Physical Sciences and
Engineering, Australian National University, Canberra ACT
0200,Australia}

\date{\today}

\begin{abstract}

We propose a scheme for scalable and universal quantum computation
using diatomic bits with conditional dipole-dipole interaction,
trapped within an optical lattice. The qubit states are encoded by
the scattering state and the bound heteronuclear molecular state
of two ultracold atoms per site. The conditional dipole-dipole
interaction appears between neighboring bits when they both occupy
the molecular state. The realization of a universal set of quantum
logic gates, which is composed of single-bit operations and a
two-bit controlled-NOT gate, is presented. The readout method is
also discussed.

\end{abstract}


\maketitle

\section{Introduction}

Quantum computers based upon the principles of quantum
superposition and entanglement are expected to provide more
powerful computation ability than classical ones in the algorithms
such as Shor's factoring \cite{Shor} and Grover's searching
\cite{Grover}. Successful implementation of quantum information
processing (QIP) would also have significant impact on many-body
quantum entanglement \cite{Q-entanglement}, precision measurements
\cite{Q-entanglement,prec-meas}, and quantum communications
\cite{commun}. To realize QIP, many schemes of quantum circuits
have been proposed including those based on trapped ions
\cite{trapped-ions}, nuclear magnetic resonance \cite{NMR}, cavity
quantum electrodynamics \cite{CQED}, linear optics \cite{LOptics},
silicon based nuclear spins \cite{solid}, quantum dots
\cite{Qdots} and Josephson junctions \cite{JJ}. Due to the long
coherence times of the atomic hyperfine states and well-developed
techniques for trapping and manipulating ultracold atoms in
optical lattices \cite{OL}, quantum computation schemes utilizing
neutral atoms become particularly attractive \cite{NA-QC,NA-QC2}.

To realize a set of universal quantum logic gates with neutral
atoms \cite{universal-QC}, the coupling between atomic bits must
be strong enough for inducing entanglement. One of the suggested
coupling mechanisms is the magnetic dipole-dipole interaction
between single atoms trapped in different sites of spin-dependent
optical lattices \cite{entanglement-Bloch}. However, due to the
very small magnetic dipole moment, one has to drive two atoms very
close together by shifting the spin-dependent optical lattice
potentials \cite{entanglement-Bloch}. If the distance between two
atomic bits is fixed and not very short, one has to induce
sufficiently large electric dipole moments with auxiliary lasers
\cite{Rydberg} or other methods. Another possibility is to use neutral
diatomic molecules with sufficiently large electric dipole moments
\cite{MQC}. However, the electric dipole-dipole interaction
between molecules can not be controllably switched off and on.
This lack of control requires additional refocusing procedures to
eliminate the effects of the non-nearest-neighboring couplings
\cite{MQC}.

Recently, applying the techniques of Raman transition, the
single-state molecules from atomic Bose-Einstein condensate
\cite{AMBEC-Heinzen}, state selective production of molecules in
optical lattices \cite{Bloch-MOL} and optical production of
ultracold heteronuclear molecules with large electric dipole
moments \cite{polar-M} have been realized successfully. These
experiments provide the potential possibility to perform quantum
computation using diatomic bits with optically induced
atom-molecular coherence. The atom-molecular coherence can also be
induced by a magnetic field Feshbach resonance
\cite{AMBEC-Wieman}.

In this article, we suggest a new scheme for quantum computation
based upon diatomic qubits with conditional electric dipole-dipole
interactions. The qubits are realized by trapping neutral Bose-condensed atoms of two different species in an optical lattice and driving the system into a Mott insulator regime with only two atoms (and only one
atom of each species) per site. Application of the well-developed technique of Raman transitions between the free atomic state and a bound molecular state at each lattice site \cite{AMBEC-Heinzen} can ensure a well-defined two-state behaviour of the diatomic system at each site, and hence the qubit states can be encoded by these two states. For certain atomic species, the ground heteronuclear
molecular state would naturally possess a large electric dipole moment. Due to the dipole-dipole interaction between
dipolar molecular states in neighbouring wells, the two-bit phase gate can be naturally
realized by free evolution. This dipole-dipole
interaction is conditional upon neighbouring qubits occupuying molecular states, and can be controllably turned on and off. Combining the
two-bit phase-gate with the single-bit Raman transitions, one can
successfully implement a set of universal gates. The
trapping and state selective production of molecules in optical
lattices \cite{Bloch-MOL} enables an excellent scalability of the processor to a lot of
qubits.

\section{Quantum computation scheme}

Let us consider two different species of Bose-condensed atoms
loaded into a one-dimensional optical lattice with the potential $V(z)=V_{0}{\text
{cos}}^{2}(kz)$, see Fig. 1 (a). If loaded adiabatically, the
atoms will occupy only the lowest Bloch band. For sufficiently
strong intensity of the laser that forms the optical lattice
potential, the tight-binding limit is reached. Under these
conditions, the system obeys the following Hamiltonian,
\begin{equation}
\begin{array}{ll}
H= & -\sum_{\left\langle i,j\right\rangle}(t_{a}a_{i}^{+}a_{j}+t_{b}b_{i}^{+}b_{j}+t_{c}c_{i}^{+}c_{j}+h.c.) \\
&+\sum_{i}\Omega_{i}(a_{i}^{+}b_{i}^{+}c_{i}+c_{i}^{+}a_{i}b_{i})+\sum_{\left\langle
i,j\right\rangle }D_{ij}n_{ci}n_{cj} \\
& +\sum_{\kappa }^{a,b,c}\sum_{i}[U_{\kappa \kappa}n_{\kappa i}(n_{\kappa i}-1)/2] \\
&+\sum_{i}(U_{ab}n_{ai}n_{bi}+U_{ac}n_{ai}n_{ci}+U_{bc}n_{bi}n_{ci}).
\end{array}\label{eq1}
\end{equation}
Here, $a_{i}^{+}$ and $b_{i}^{+}$ ($a_{i}$ and $b_{i}$) are
bosonic creation (annihilation) operators for atoms on site $i$,
$c_{i}^{+}$ ($c_{i}$) are corresponding operators for molecules on
site $i$, and $n_{\kappa i}=\kappa_{i}^{+}\kappa_{i}$ with
($\kappa=a, b, c$) are particle numbers. The symbol ${\left\langle
i,j\right\rangle }$ represents summing over the nearest-neighbors
and $h.c.$ denotes the Hermitian conjugate terms. The first term
describes the tunneling between neighboring sites with the tunneling
strength $t_{\kappa}$. The second term corresponds to the coupling
between atoms and molecules with Rabi frequencies $\Omega_{i}$.
The third term is the electric dipole-dipole interaction between
molecules with the coefficients $D_{ij}$ determined by the dipole
moments and the lattice spacing. The last two terms describe the
inter- and intra-component scattering with the coefficients $U_{\kappa
\kappa^{'}}$ determined by the s-wave scattering lengths.
\begin{figure}[h]
\rotatebox{0}{\resizebox *{9.0cm}{8cm} {\includegraphics
{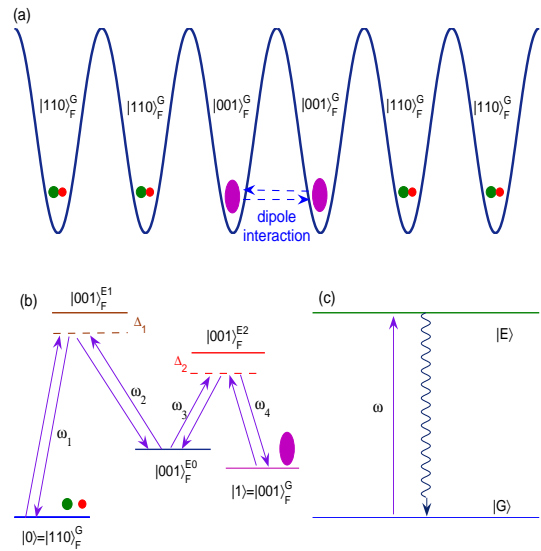}}} \caption{\label{fig:epsart} (Color online) Scheme
of quantum computation using diatomic qubits with conditional
dipole-dipole interaction. (a) Diatomic qubits in one-dimensional
optical lattices. The dipole-dipole interaction appears when
neighboring bits occupy molecular states. (b) Single-bit
operation with optimally controlled processes sandwiched Raman
transition (see text). (c) Read-out with photon scattering (see
text).}
\end{figure}

\subsection{Initialization}

To initialize the processor, one can ramp up the potential depth
after the two species of ultracold atoms are loaded into the
optical lattice. For a sufficiently deep potential, the
Mott insulator phase with $|n_{ai}=1, n_{bi}=1, n_{ci}=0 \rangle
_{F}^{G}$ for every site can be easily obtained \cite{MI}. Here, $F$ denotes the Fock states, and $G$ denotes the
ground states. Raman pulses can coherently couple the trapped atoms in the scattering state $|110\rangle_{F}^{G}$ and the diatomic heteronuclear molecular ste $|001\rangle_{F}^{G}$ at each site, which can therefore encode the qubit states $|0\rangle$ and $|1\rangle$, respectively. The Mott insulator state in the absence of coupling fields corresponds to the
qubits state $|000...\rangle$.

\subsection{Universal set of quantum logic gates}

By properly choosing the atomic species, the heteronuclear
molecules, such as RbCs and KRb \cite{polar-M}, appear with very
large electric dipole moments. By combining the techniques of
coherent Raman transition and optimally controlled process (OCP)
\cite{OCP}, the limit of Franck-Condon principle can be overcome.
The single-bit operations (i.e., preparation of an arbitrary superposition
of the atomic state $|110\rangle_{F}^{G}$ and the ground molecular
state $|001\rangle_{F}^{G}$) can be realized with a Raman pulse
sandwiched by two OCPs, see Fig. 1 (b). The first OCP transfers
the ground molecular state to an excited one, the Raman pulse
realizes the required superposition of the excited molecular state
and the unbounded state of atoms, and then the second OCP
transfers the excited molecular state back to the ground one.

The core task of quantum computation is to realize a set of universal
quantum logic gates, such as single-bit operations combined with
two-bit controlled-NOT gates \cite{universal-QC}. As shown in Fig.
1 (b), the single-bit operations can be performed with optical
stimulated Raman processes. A $R_{y}(\pi)$ pulse will transfer
$|0\rangle$ (or $|1\rangle$) to $|1\rangle$ (or $|0\rangle$), and
a $R_{y}(\pi/2)$ pulse will transfer $|0\rangle$ (or $|1\rangle$)
to $\frac{1}{\sqrt{2}}(|0\rangle+|1\rangle)$ [or
$\frac{1}{\sqrt{2}}(-|0\rangle+|1\rangle)$]. When all laser
frequencies are detuned far from the transition frequencies to the
excited molecular state, the excited molecular states will not be
populated.

Because of the short distance (an order of a wavelength in an
optical lattice) between neighboring bits and the same transition
frequency for all bits, it is very difficult to selectively
address a particular qubit by focusing the laser beams only on a
particular site. Fortunately, similar to the well-developed
techniques of gradient magnetic field in nuclear magnetic
resonance, the transition frequencies for different bits can be
distinguished by applying an external electric field \cite{MQC},
\begin{equation}
\overset{\rightharpoonup}{E}_{ext}=\left({E_{0}+z\frac{dE}{dz}}\right)\overset{\rightharpoonup}{e}_{x}
=(E_{0}+gz)\overset{\rightharpoonup}{e}_{x},
\end{equation}
in the direction $\overset{\rightharpoonup}{e}_{x}$ perpendicular
to the lattice direction $\overset{\rightharpoonup}{e}_{z}$, with
a gradient $g$ along the lattice direction
$\overset{\rightharpoonup}{e}_{z}$. To dominate the system, the
external electric field must satisfy the condition,
\begin{equation}
\text{Min}(\left|{\overset{\rightharpoonup }{E}_{ext}}\right|)\gg
\left|{\overset{\rightharpoonup
}{E^{i}}_{int}}\right|=\left|{\underset{j\neq i}{\sum
}\frac{-\overset{\rightharpoonup }{d}_{j}n_{cj}}{4\pi \epsilon
_{0}|r(j-i)|^{3}}}\right|.
\end{equation}
Here,  $\overset{\rightharpoonup }{E^{i}}_{int}$ is the internal
electric field on site $i$ created by the molecules in the
neighboring site, $\overset{\rightharpoonup }{d}_{j}$ is the electric
dipole moment for a single molecule on the $j-$th site, $r$ is the
distance between two nearest-neighboring sites, and the molecular
occupation numbers $n_{cj}$ are either $0$ or $1$. The difference between
transition frequencies of nearest-neighbor bits,
\begin{equation}
\Delta\nu = \frac{\Delta E_{ext}d}{\hbar}=\frac{gdr}{\hbar},
\end{equation}
increases with the gradient. Thus, for a sufficiently large
gradient, the selective addressing can be implemented by properly
choosing frequencies of the laser fields. In Table I, we show
$\Delta\nu$ for different diatomic bits $XY$ ($X$ = Li, Na, K and
Rb; $Y$ = Na, K, Rb and Cs) with $g=1.0$ $V/cm^{2}$ and $r=420$
$nm$ corresponding to the optical lattices formed by a laser with
wavelength $\lambda=840$ $nm$ \cite{Bloch-MOL}. All $\Delta\nu$
are in order of $100$ Hz which are large enough to guarantee
selective addressing a particular qubit without changing its
neighbors.

\begin{flushleft}
Table I. Difference between transition frequencies of nearest-neighbor bits with $g=1.0$ $V/cm^{2}$ and $r=420$ $nm$. The
related values for electric dipole moments are obtained from
\cite{EDM}.
\end{flushleft}
\begin{tabular}{r|rrrr}
$\Delta \upsilon (XY)$ & Na & K & Rb & Cs \\ \hline
Li & 70.41 Hz & 464.97 Hz & 548.66 Hz & 728.00 Hz \\
Na &  & 365.33 Hz & 442.38 Hz & 611.10 Hz \\
K &  &  & 85.02 Hz & 255.07 Hz \\
Rb &  &  &  & 167.39 Hz
\end{tabular}
\vskip 0.5cm

To implement two-bit gates, one has to switch on the conditional
dipole-dipole interaction between molecular states of neighboring
bits,
\begin{equation}
D_{ij}=\frac{1}{4\pi \epsilon _{0}}\times \frac{\stackrel{\rightharpoonup }{%
d}_{i}\cdot \stackrel{\rightharpoonup }{d}_{j}}{|r(j-i)|^{3}}.
\end{equation}
In this formula, we have assumed that both dipole moments are
oriented along the external electric field. Because of the
dominant strength of $\overset{\rightharpoonup }{E}_{ext}$, the
electric dipole moments for the molecular ground state in
different lattice sites have the same direction. In contrast to
the quantum computation schemes utilizing polar molecules \cite{MQC}, the
non-nearest-neighbor interactions can be switched off locally by
transferring the non-nearest-neighbor bits into free atomic
states. That is, the conditional dipole-dipole interaction
$D_{ij}n_{ci}n_{cj}$ is switched off when the molecular occupation
numbers $n_{ci}$ or $n_{cj}$ equal to zero. The controllability of
these dipole-dipole interactions removes the need for the
refocusing procedure \cite{RefocusingP} which eliminates the
effects of non-nearest-neighbor interactions \cite{NMR,MQC}.

Now let us analyze the realization of two-bit phase gates
according to the dynamics governed by the Hamiltonian (\ref{eq1}) with
parameters in deeply insulating region of two different atoms or a
molecule per site. Due to the dipole-dipole interaction only
existing between molecular states, in free evolution the quantum
logic state $|11\rangle$ will naturally acquire a phase shift.
That is, an arbitrary two-bit state will be transformed as follows:
\begin{equation}
\begin{array}{ll}
C_{00}\left| 00\right\rangle +C_{01}\left| 01\right\rangle
+C_{10}\left| 10\right\rangle +C_{11}\left| 11\right\rangle
\\
\longrightarrow C_{00}\left|00\right\rangle +C_{01}\left|
01\right\rangle +C_{10}\left| 10\right\rangle +C_{11}\exp
(i\varphi )\left| 11\right\rangle,
\end{array}
\end{equation}
with the phase shift,
\begin{equation}
\varphi=\frac{D_{12}t} {\hbar}= \frac{d_{1}d_{2}t} {4\pi
\epsilon_{0} \hbar r^{3}},
\end{equation}
determined by the coupling strength $D_{12}$ and the evolution
time $t$. Here the coefficients $C_{ij}$ ($i,j=0,1$) denote the
probability amplitudes, and $d_{1,2}$ are electric dipole
moments.

With the phase gate, it is easy to prepare four Bell states and
construct a controlled-NOT gate \cite{QC-Book}. The Bell state
$\frac{1}{\sqrt{2}}(|00\rangle+|11\rangle)$ can be prepared from
the initialized state $|00\rangle$ (Mott insulating phase with two
atoms per site) with the following steps: (i) a two-bit
$\frac{\pi}{2}$ pulse, the initialized state is transferred into
$\frac{1}{2}(|00\rangle + |01\rangle + |10\rangle + |11\rangle)$;
(ii) a $\pi$ phase gate, the state freely evolves to
$\frac{1}{2}(|00\rangle + |01\rangle + |10\rangle +
\text{exp}(i\pi)|11\rangle)$; and (iii) a single-bit
$\frac{\pi}{2}$ pulse for the first qubit. The other three Bell
states can be obtained from this state by free evolution ($\pi$
phase gate) or single-bit operation (single-bit $\pi$ pulse). It
is well known that the controlled-NOT gate can be constructed by
two target-bit Hadamard gates sandwiching a $\pi$ phase gate
\cite{QC-Book}. Usually, to simplify the pulse sequences, the
first Hadamard gate is replaced with a single-bit
$R_{y}(-\frac{\pi}{2})$ pulse and the second one is replaced with
a single-bit $R_{y}(\frac{\pi}{2})$ pulse. This means that the
time for a controlled-NOT gate equals the time for a $\pi$ phase
gate plus the time for a single-bit $2\pi$ pulse. Due to the very
short time for a single-bit $2\pi$ pulse at large Rabi frequency,
the total time for a controlled-NOT gate is dominated by the time
for a $\pi$ phase gate. By choosing the same parameters as for
Table I, and ignoring the short times for single-bit operations,
we can estimate the possible numbers of controlled-NOT gates per
second,
\begin{equation}
N= \frac{D_{12}} {\hbar\pi} = \frac{d_{1}d_{2}} {4\pi^{2}
\epsilon_{0} \hbar r^{3}}.
\end{equation}
The values of $N$ for different diatomic bits $XY$ are shown in
Table II. Most of the $N$ values are of the order of $10^{4}$,
which guarantees that the system can successfully implement a lot
of quantum logic gates before it loses quantum coherence.

\begin{flushleft}
Table II. Possible numbers of controlled-NOT gates per second.
\end{flushleft}
\begin{tabular}{r|rrrr}
$N(XY)$ & Na & K & Rb & Cs \\ \hline Li & $1.14\times 10^{3}$ &
$4.99\times 10^{4}$ & $6.94\times 10^{4}$ & $1.22\times 10^{5}$ \\
Na &  & $3.08\times 10^{4}$ & $4.51\times 10^{4}$ & $8.62\times 10^{4}$ \\
K &  &  & $1.66\times 10^{3}$ & $1.50\times 10^{4}$ \\
Rb &  &  &  & $6.46\times 10^{3}$%
\end{tabular}
\vskip 0.5cm

\subsection{Readout}

There are two different choices for reading out the final states.
The first one is photon scattering which has been used to detect
states of ion trap quantum computer \cite{photon-scattering}. The
basic idea is illuminating the diatomic qubits with a circularly
polarized laser beam tuned to the cycling transition from the
ground state $|G\rangle$ of the selected particle (atom A, atom B,
or molecule C) to the corresponding excited state $|E\rangle$, see
Fig. 1 (c). If there are particles in $|G\rangle$, the
photomultiplier will detect the scattered photons. Otherwise,
there are no scattered photons. The second one is state-selective
resonant ionization \cite{MQC,ionization}. In this method, one can
apply a resonant laser pulse to selectively ionize the molecular
ground state (qubit state $|1\rangle$) after rapidly switching off
the external gradient electric field. Then the electrons and ions
can be detected by imaging techniques.

\subsection{Open problems}

In real experiments, many practical factors must be taken into
account. One is the strength of the optical lattice potential
needed to keep the system in the Mott insulating phase with two
different atoms or a heteronuclear molecule per site. In the further study, it would be
interesting to analyze the details of quantum phase transitions to
quantify the parameter region for the insulating phase, in
particular, the effects of conditional dipole-dipole interaction
and Raman coupling between atomic and molecular states. Another
important factor is decoherence. As pointed out in previous
studies \cite{NA-QC}, the decoherence from spontaneous emission
can be avoided by choosing lasers far detuned from atomic
transitions to form the optical lattices. In our model, we have also neglected
the motional states localized in each lattice site. To avoid the
coupling between motional excitation and gate operation, similar
to the proposal by Jaksch et al. \cite{NA-QC2}, one has to confine
the qubits in deep Lamb-Dicke regimes to eliminate the significant
momentum transfer to the qubits from the operational lasers.
However, some vibrational and rotational molecular states and even
some hyperfine states may be excited by the Raman processes. The
effects of these excited states will bring a source of decoherence
which is not easy to eliminate.

\section{Summary and discussion}

In conclusion, we have demonstrated the possibility of using
diatomic bits with conditional dipole-dipole interaction to
implement scalable and universal quantum computation. By trapping
the ultracold diatomic bits within optical lattices, the system can be
scaled to a large number of qubits. Combination of the coherent
Raman transition between atomic and heteronuclear molecular states
with the free evolution involving conditional dipole-dipole
interaction makes the QIP based upon these diatomic qubits
universal. Unlike the previous proposals for quantum computation
in optical lattices, our proposal does not require relative
shifting of the spin-dependent optical lattice potentials
\cite{NA-QC,NA-QC2}, coupling to Rydberg states with large
electric dipole moments \cite{NA-QC,Rydberg} or refocusing
procedures to eliminate the effects of non-nearest-neighbor
interaction \cite{MQC}. We have also shown that the selective
addressing of qubits can be realized by applying an external
gradient electric field, and that the strength of dipole-dipole
interactions guarantees the performance of a large number of
quantum logic gates (in order of $10^{4}$) per second.

Our analysis can also be applied to the case of two
different kinds of Fermi atoms in optical lattices. For the system
of Fermi atoms, due to the Pauli blocking, the s-wave scattering
between Fermi atoms of the same kind is absent. That is, the
Hamiltonian (\ref{eq1}) has no terms containing $U_{aa}$ or
$U_{bb}$.

\acknowledgments The authors acknowledge discussions with Yuri S.
Kivshar. This work is supported by the Australian Research Council
(ARC).

\end{document}